\documentstyle[prl,twocolumn,aps]{revtex}
\input{epsf}
\begin{document}
\draft
\twocolumn[\hsize\textwidth\columnwidth\hsize\csname @twocolumnfalse\endcsname
\title{Suppression of Ground-State Magnetization in Finite-Sized
Systems Due to Off-Diagonal Interaction Fluctuations}\vspace*{-0.22cm}
\author{Philippe Jacquod and A. Douglas Stone}

\address {Department of Applied Physics, P. O. Box 208284, Yale University,
New Haven, CT 06520-8284}

\date{\today}

\maketitle
\begin{abstract}
We study a generic model of interacting fermions in a finite-sized disordered
system. We show that the off-diagonal interaction matrix elements induce
density of states fluctuations which generically favor a minimum spin ground
state at large interaction
amplitude, $U$.  This effect competes with the exchange effect which favors
large magnetization at large $U$, and it suppresses this
exchange magnetization
in a large parameter range.  When off-diagonal fluctuations dominate, the
model predicts a spin gap which is larger for odd-spin ground states
as for even-spin, suggesting a simple
experimental signature of this off-diagonal effect in Coulomb blockade
transport measurements.
\end{abstract}
\pacs{PACS numbers : 73.23.-b, 71.10.-w, 71.24.+q, 75.10.Lp}
]

Ferromagnetic instabilities result from the combined
effect of the electronic interactions together with the Pauli
principle. The interaction energy can be minimized when the
fermionic antisymmetry requirement is satisfied by the spatial
wavefunction leading to alignment of spins and a large ground-state spin
magnetization (a familiar example of this is Hund's rule in atoms). In
contrast when the interaction is
weak minimal spin is favored because it costs kinetic energy to flip a
spin as it must then be promoted to a higher energy level.  When treating
ferromagnetism in metals,
because of the locality of the Pauli principle, magnetic
instabilities are usually studied in the framework of the Hubbard
model
taking only the short-range part of electronic interactions into
account,
so that only pairs of electrons of opposite spin interact. As the
magnetization increases
the number of interacting pairs decreases and the system spontaneously
magnetizes
for sufficiently strong interactions.
In a finite-sized system this {\it Stoner instability}\cite{stoner}
occurs when the typical
exchange interaction between two states close to the Fermi energy
is equal to the one-particle level spacing which for a clean system
with Hubbard interaction gives 
$U_c=\Delta$.

There has been much recent interest in the Stoner instability of
finite-size
disordered metals such as quantum dots and metallic nanoparticles
\cite{nano}.
Building on earlier perturbative work \cite{altshuler}
Andreev and Kamenev \cite{andreev} recently found a significant
reduction of the Stoner
threshold in disordered systems due to spatial correlations in
diffusive
wavefunctions which enhance the average exchange term.
More recently, Brouwer et. al. \cite{brouwer} considered the effect of
mesoscopic wavefunction fluctuations
and found an associated increase of the probability
of non-zero ground state spin magnetization below the Stoner
threshold. Within a similar model, Baranger et. al.\cite{baranger}
proposed that spontanous magnetization effects could explain kinks in
the field-dependence
of Coulomb blockade resonances.
The purpose of the present paper is to point out a competing effect of
interactions
which {\it suppresses} the probability of ground-state magnetization
and
has not been treated in any of the previous works on
itinerant magnetism of
disordered systems. The mean-field treatments leading to an exchange
term
in the effective hamiltonian for disordered metals neglects the
effects
of {\it off-diagonal} interaction matrix elements \cite{andreev};
however
it is well-known from studies of nuclei and atoms \cite{brody} that
the
band-width of the many-body density of states in finite interacting
fermi
systems is actually {\it determined} by the fluctuations of these
off-diagonal
matrix elements.  When one introduces the spin degree of freedom into
these 
models we shall see below that 
one immediately finds that these fluctuations are largest for the
states
of minimal spin.  This effect then strongly increases the probability
that
the extremal (low-lying) states in the band are those of minimal spin
and
opposes the exchange effect.  We expect this effect to be significant
in
quantum dots and to suppress the possibility of high spin ground
states.

We start from the Hamiltonian for $n$ spin-$1/2$ particles

\begin{eqnarray}\label{hamiltonian}
H & = & \sum \epsilon_{\alpha}
c^{\dagger}_{\alpha,s} c_{\alpha,s}
+\sum
  U_{\alpha,\beta}^{\gamma,\delta}
c^{\dagger}_{\alpha,s} c^{\dagger}_{\beta,s'} c_{\delta,s'}
c_{\gamma,s}
\end{eqnarray}

\noindent $s^{(')} = \uparrow,\downarrow$ are spin indices.
The $m/2$ different one-body energies are distributed as
$\epsilon_{\alpha} \in [-m/2;m/2]$ so as to fix $\Delta \equiv 1$ with
spin degeneracy. The interaction
commutes with the $z$-component $\sigma_z$ of the total magnetization
$\sigma$
so that the Hamiltonian acquires a block structure where blocks are 
labelled by $\sigma_z$ and due to Spin Rotational Symmetry (SRS)
subblocks 
of given $\sigma \ge \left| \sigma_z \right|$ appear within each of 
these blocks. Each block's size is given in term of
binomial coefficients as $N(\sigma_z) =
\left(_{n/2-\sigma_z}^{m/2}\right)
\left(_{n/2+\sigma_z}^{m/2}\right)$. 

The Hamiltonian (1) can be viewed as a generic model of interacting
fermions expressed in the basis of Slater determinants constructed
from
the eigenstates $\psi_{\alpha}$ of the corresponding free fermions
model
$H_0 \equiv \sum \epsilon_{\alpha}
c^{\dagger}_{\alpha,s} c_{\alpha,s}$.
In this basis the interaction matrix elements are given by
$ U_{\alpha,\beta}^{\gamma,\delta} = \int d\vec{r} d\vec{r'}
U(\vec{r}-\vec{r'}) \psi_{\alpha}^*(\vec{r})  \psi_{\beta}^*(\vec{r'})
  \psi_{\gamma}(\vec{r})  \psi_{\delta}(\vec{r'})$, where
$U(\vec{r}-\vec{r'})$ is the interaction potential.
Due to disorder or chaotic boundary scattering the wavefunctions
$\psi_{\alpha}$ have a random character leading to fluctuations
in $ U_{\alpha,\beta}^{\gamma,\delta}$ around their average value.
We take these fluctuations
to be random with a zero-centered gaussian distribution of width $U$.
This gives a contribution $\bar{H}$ similar to the second term in the 
right-hand side of (\ref{hamiltonian}) with
a distribution $ P(U_{\alpha,\beta}^{\gamma,\delta}) \propto
e^{-(U_{\alpha,\beta}^{\gamma,\delta})^2/2U^2}$ of interaction matrix
elements.
Only diagonal matrix elements $U_{\alpha,\beta}^{\alpha,\beta}$ and
$U_{\alpha,\beta}^{\beta,\alpha}$ have a nonzero average
leading to mean-field charge-charge and spin-spin
diagonal interactions \cite{andreev}.
We neglect the charge-charge contribution as it has
no influence on the magnetization and
this leaves us with the following effective Hamiltonian

\begin{equation}\label{effhamiltonian}
{\cal H} = H_0+\bar{H} - \lambda U \sum \vec{s}_{\alpha}
\vec{s}_{\beta}
\end{equation}

\noindent $\vec{s}_{\alpha} \equiv \sum_{s,t}
c^{\dagger}_{\alpha,s} \vec{\sigma}_{s,t} c_{\alpha,t}$ are spin
operators
and the ferromagnetic spin-spin interaction has a strength $\lambda
U>0$.
Without it ($\lambda=0$), the Hamiltonian (\ref{effhamiltonian})
within each spin block is precisely the Two-Body Random Interaction
Model
(TBRIM) introduced in nuclear physics \cite{french}, and used
to study thermalization \cite{flam} and the emergence of quantum chaos
in few-body systems \cite{m2body} and statistical features observed in
shell model calculations \cite{flam2,zel}.
A similar model has been shown recently to be consistent with the
observed gaussian distribution of peak-spacings
in Coulomb blockade resonances through quantum dots \cite{yoram}.
One key feature of the TBRIM is that the many-body density of states
(MBDOS) has an approximately gaussian shape with a variance
proportional to the {\it connectivity } $K$, i.e. the number of
non-zero matrix elements in each row \cite{french}.
Similarly we can estimate the variance of the MBDOS 
of the Hamiltonian (\ref{effhamiltonian}) for fixed $(\sigma,\sigma_z)$
and 
$U/\Delta \gg 1$ as

\begin{equation}\label{dosvar}
\frac{1}{N(\sigma_z)} \sum_{I,J} \bar{H}_{I,J}^2 
\delta(\sigma_z^{(I)}-\sigma_z) \delta(\sigma^{(I)}-\sigma)
\approx K U^2
\end{equation}

\noindent where $ \bar{H}_{I,J} = \langle I| \bar{H} |J \rangle $ and
$|I \rangle$ refers to a Slater determinant.
Hence each block's bandwidth goes as $\sqrt{K} U$ with a
$\sigma_z$-dependent connectivity which can be expressed
as $K(n,m,\sigma_z) \approx 1+C(n/2+\sigma_z,m/2)+C(n/2-\sigma_z,m/2)
+1/2 ((n/2)^2-\sigma_z^2)((m/2-n/2)^2-\sigma_z^2)$ in term of the
function
$C(n,m)=n(m-n)+n(n-1)(m-n)(m-n-1)/4$. The factor $1/2$ in front of the
last contribution to $K$ is needed to take into account the effect of 
SRS.  In the dilute limit approximately half of the spin-flip
transitions
which conserve $\sigma_z$ are not allowed because they don't  conserve
total
spin (e.g. change a singlet to triplet).
The estimate (\ref{dosvar}) assumes that each matrix element has the 
same variance, which for a generic off-diagonal element is $\sim U^2$.
However it is easily seen that
diagonal matrix elements $\bar{H}_{I,I}$ have an enhanced variance $
\sim 
(3n^2/4+\sigma_z^2)^2 U^2$  which induces deviations
from (\ref{dosvar}) for large filling and large magnetization. 
Nevertheless, these matrix elements can be neglected in the dilute and 
weakly-magnetized  limit $1 \ll n/2+\sigma_z \ll m$ where 
the larger number of off-diagonal matrix elements 
dominates the variance when $(m-n)^2 \gg (3n^2/4+\sigma_z^2)$.
In the right inset to Fig. 1 we show plots of $K(n,m,\sigma_z)$
for different filling factors $\nu=n/m$ as well as a comparison with
the true
variance of the MBDOS for $\nu=3/8$. Deviations from
the estimate (\ref{dosvar}) are small, even at this
rather large filling and increase with increasing magnetization in
agreement
with the above reasoning. Moreover in the left inset to Fig. 1 we 
show that the full MBDOS follows the
scaling (\ref{dosvar}) with significant deviations only at large
magnetization, so that it is plausible that this scaling also 
determines the tails in which the ground state energies will be found.

\begin{figure}
\epsfxsize=3.3in
\epsfysize=2.6in
\epsffile{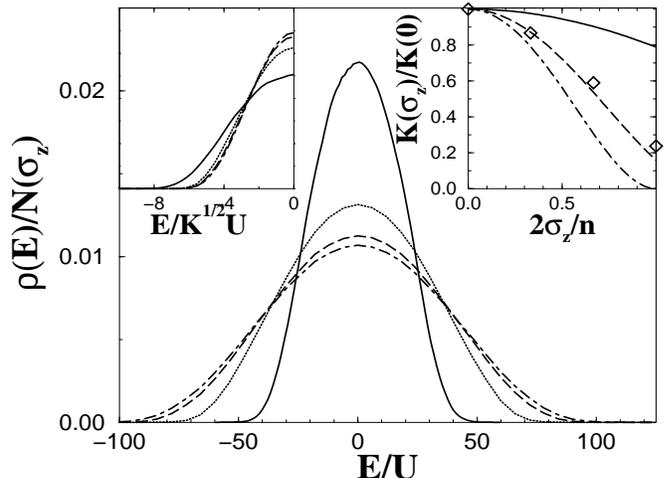}
\caption{Density of states for the Hamiltonian $\bar{H}$
with $n=6$ particles and $m=16$ orbitals,
corresponding to the magnetization blocks $\sigma_z=-3$ (solid line),
-2 (dotted line), -1 (dashed line) and 0 (dotted-dashed line). Left
inset:
rescaled density of states showing the approximate scaling in
$E/K^{1/2}U$.
Right inset :
Normalized connectivity $K(\sigma_z)/K(0)$ vs. magnetization
for filling factors $\nu=1/10$ (solid line), 3/8 (dashed line) and 1/2
(dotted-dashed line). We compare this estimate to the true variance
for
$\nu=3/8$ obtained numerically from the data of the main figure
(diamonds).}
\label{fig:dos}
\end{figure}

Our essential finding follows from the simple features of the model
already
stated. The full MBDOS is a sum of approximately gaussian
contributions from each
spin block with a variance proportional to the corresponding
connectivity.
The latter is a {\it monotonously decreasing} function of $\sigma_z$.
Hence the broadest MBDOS corresponds to the minimally magnetized block
and the ground state will be found in this block with increased
probability
\cite{srs}. Assuming, as just discussed, that the tails of the
distribution 
scale with the variance with a factor $\beta$ and
neglecting contributions arising from $H_0$, the typical
spin gap can be estimated (for $\lambda = 0$) as

\begin{equation}\label{spingap}
\Delta_s^{U} \approx \beta U[ \sqrt{K(|\sigma_{min}|)} -
\sqrt{K(|\sigma_{min}| + 1)}]
\end{equation}

\noindent This multiple gaussian structure of the MBDOS and the scaling with
$\sqrt{K}$ obtained from numerical calculations are shown in Fig. 1
for $\lambda =0$. For $\lambda \ne 0$, the
spin-spin interaction induces relative shifts of each block's MBDOS
which eventually will shift the finite spin blocks sufficiently to 
overcome the larger fluctuations of the minimal
spin MBDOS. This is the competition between exchange and off-diagonal
fluctuations already mentioned.  However for reasonable values of
$\lambda$ the off-diagonal fluctuations strongly reduce the
probability
of exchange-induced magnetization (see Fig. 3).

\begin{figure}
\epsfxsize=3.3in
\epsfysize=2.6in
\epsffile{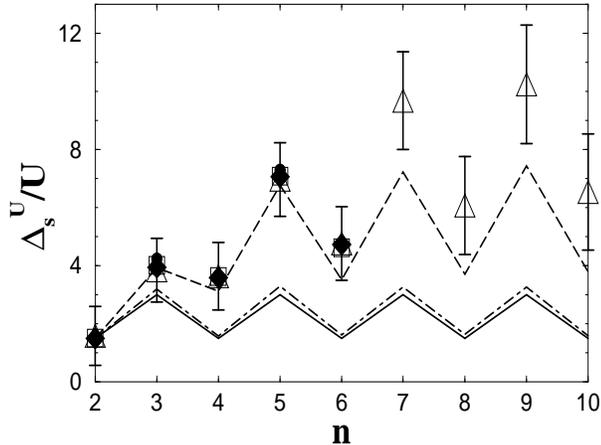}
\caption{Dependence of the finite-size spin gap in the number $n$ of
particles.
Points correspond to numerical results for $m=10$ (full circles), 12
(empty
squares), 14 (full diamonds) and 16 (empty triangles) and the solid
line to the
dilute estimate (\ref{spingap}) with a numerical factor $\beta=1.5$.
For the case $m=16$ and 1000 Hamiltonian realizations,
the error bars indicate the r.m.s. of the gap distribution while the
dashed 
and dotted-dashed lines show the numerically computed variances 
(Left-hand side of eq. (3)) for the full Hamiltonian and after setting
to zero non-generic interaction matrix elements respectively.}
\label{fig:gapinf}
\end{figure}

In Fig. 2 we show the computed spin gap
$\Delta_s^{U}$ between the minimally magnetized ground-state and the
first
spin excited level for $\lambda=0$
in the limit of dominant interaction, i.e. neglecting
$H_0$ in (\ref{hamiltonian}). One of the main features
is a strong even-odd effect which is reminiscent of a similar behavior
in the limit of vanishing interactions. 
However the origin here is the fluctuating interaction and the 
energy differences scale as $U$ instead of $\Delta$. 
We next note that the gap first increases with increasing
number of particles before it seems to stabilize above $n=6$. We 
have checked (dashed and dot-dashed lined in Fig. 2) 
that this behaviour, which is not captured by the dilute
estimate (\ref{spingap}), is partly due to the
neglect in (\ref{dosvar}) of nongeneric matrix elements with enhanced 
variance mentioned above. However,
even though the exact variance gives a much better estimate, it
still underestimates the gap at larger $n$ and we have numerically
determined 
that this is due to a strong positive correlation of the 
ground state energies in adjoining spin blocks ($\sim 0.9$). Such
correlations, 
although interesting, are not suprising since the different
block hamiltonians are not statistically independent (many of the same
two-body
matrix elements appear in both).

We next switch on the mean-field spin-spin interaction $\lambda >0$
which induces energy shifts of 
$-\lambda U \left|\sigma_z\right|(\left|\sigma_z\right|+1)$
\cite{srs}.
On average the spin gap becomes $\Delta_s = \Delta_s^{U} -
\bar{\lambda}U$,
where $\bar{\lambda}=(5-(-1)^n)\lambda/2$, i.e.
the relative shift between the two lowest magnetized blocks
is larger for odd number of particles. The variance of the
gap distribution is unaffected and we can already conclude that the
probability $P(\sigma_z>0)$ of finding a magnetized ground-state
is reduced by the off-diagonal matrix elements and saturates above a
finite value $U_c$, since
the width of the gap distribution is proportional to its average $\sim
U$.
This is shown on Fig. 3 where $P(\sigma_z>0)$ is plotted against
$U/\Delta$ for different values of $\lambda$. On the same graph we
show
numerical results obtained after setting to zero the off-diagonal 
matrix elements. The data unambiguously reflect the strong
demagnetizing influence of the off-diagonal matrix elements.

\begin{figure}
\epsfxsize=3.3in
\epsfysize=2.4in
\epsffile{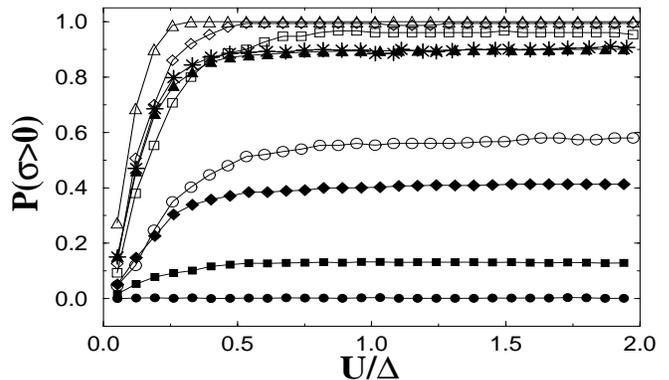}
\caption{Probability for a magnetized ground-state as a function of
$U/\Delta$ for 2000 realizations of
Hamiltonian (\ref{effhamiltonian}) with $n=5$ and $m=10$,
$\lambda = 2,4,5$ and $7$ (full symbols, from bottom to top).
Empty symbols show the corresponding
curves after setting to zero the off-diagonal matrix elements.
Stars correspond to $n=5$, $m=14$ and $\lambda=7$. }
\label{fig:prob}
\end{figure}

The physically
relevant value of $\lambda$ will depend on the microscopic
details of the system. Indeed $\lambda$ is given by half the ratio of
the mean exchange interaction with the fluctuations of off-diagonal
matrix elements \cite{andreev}
$\lambda = \langle
U_{\alpha,\beta}^{\beta,\alpha} \rangle/2 \
{\rm r.m.s.}(U^{\gamma,\delta}_{\alpha,\beta})$.
Semiconductor quantum dots with poor screening and extended,
chaotic single-particle
wavefunctions should have $\lambda \approx 1$ while
for extended diffusive metallic systems, we get a much larger
spin-spin interaction $\lambda \sim g$
where $g > 1$ is the system's conductance \cite{altshuler}.
From Fig. 3 however, we see that even in the regime of
dominating interactions, a nonmagnetized ground state is more probable
for $\lambda \ll 5$ ($\bar{\lambda} U \ll \Delta_s^{U}$).

We finally consider the influence of an external magnetic field
which only introduces a Zeemann coupling. This situation can be
experimentally realized by applying a magnetic field in the plane of a
two-dimensional electron gas. This Zeemann term does not
affect the Hamiltonian's block structure, but only shifts
each block's MBDOS by an amount $g \mu_B B \sigma_z$. Due to the
spin gap discussed above, a finite magnetic field of average magnitude
$\langle B_c \rangle = \Delta_s/g \mu_B$ is necessary to
magnetize the system. The even-odd effect emphasized
in Fig. 2 results in a critical field to flip one spin which is
significantly larger for a $\sigma_z = \pm 1/2$ (odd) ground state as
compared to a $\sigma_z = 0$ (even) ground state.  More generally
the gap for a lower spin state is smaller than that for the state of
higher spin (this follows from the right inset to Fig. 1). This aspect
of our theory can be tested experimentally by studying the
in-plane magnetic field dependence of
the position of Coulomb blockade conductance peaks at very low
temperature
$T \ll \Delta$. The resonant gate voltage is given by a difference of
two
many-body ground-state energies
$e V_g^n = E^0_{n+1} - E^0_{n}$, and it is always the difference of
an even-odd pair. The peak position behaves like

\begin{eqnarray}
e V_g^n(B) & = &
E^0_{n+1} - E^0_{n} + g \mu_B B \delta \sigma_z(n)
\end{eqnarray}

\noindent where $\delta \sigma_z(n)$ is the magnetization difference
between the two consecutive ground-states. Without magnetization
$\delta \sigma_z(n)= (-1)^n/2$ and one has
$|\partial V_g/\partial B| = g \mu_B/2$. As $B$ is increased the
ground-state
with even number of electrons is most likely to
magnetize first, exactly reversing the slope of two consecutive peaks;
then as the field increases further the odd state will likely flip,
restoring the original slope.  As long as consecutive ground states
never differ by more than one unit of spin the absolute
value of the slope will remain constant as the system polarizes.

\begin{figure}
\epsfxsize=3.3in
\epsfysize=2.1in
\epsffile{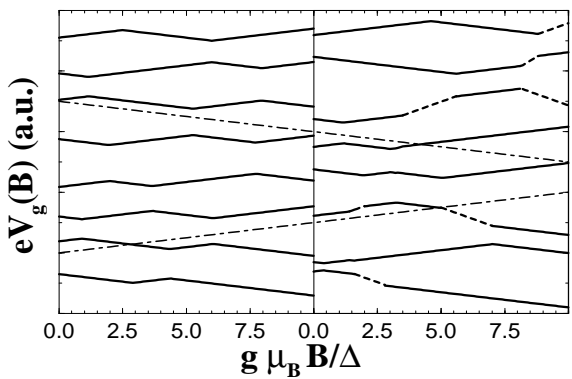}
\caption{Schematic of the conductance peaks in a 2D
quantum dot as a function of an in-plane magnetic field for $m=14$,
$\lambda = 1$ (left) and 2.5 (right) and $U/\Delta=4$ corresponding
to the addition of the
$n=3,4,...10^{\rm th}$ electrons (from bottom to top). 
Dotted-dashed lines indicate slopes of $\pm g \mu_B/2$. Larger
slopes for which spin-blockade effects strongly reduce the peak 
height [15] are indicated by dashed segments.
Note that due to the subtraction of the average
charge-charge interaction, the model does not reproduce the charging
energy so that the vertical distance between consecutive peaks is
arbitrary.}
\label{fig:peaks}
\end{figure}

However if there exist many magnetized ground states
then one expects a range of slopes to occur.
In this case the corresponding peak heights will be strongly reduced
by
the spin blockade mechanism (see dashed lines on 
Fig. 4) \cite{weinmann} which should be easily visible experimentally.
This argument neglects changes in the $g$-factor of the
electron with
changing $n$, which presumably are slow.
The even-odd behavior of $\Delta_s$ is qualitatively similar
to the noninteracting case, however the scale in $g \mu_B B$
over which spin flips occur is determined by $U$ and not $\Delta$.
Typically, this results in an increase of the field necessary 
to achieve full polarization on the dot. This is illustrated on Fig. 4
where
the peaks positions are drawn as a function of the
Zeemann coupling for $\lambda=1.5$ and 3. It is clearly seen that at
small $\lambda$, $|\partial V_g/\partial B|$ is constant and
corresponds to a minimal $\delta \sigma_z$,
while increasing $\lambda$ gives different slopes in agreement with
the above reasoning.

Numerical computations were performed at the Swiss Center
for Scientific Computing and at the National Energy Research
Scientific
Computing Center. Work supported by the NSF grant PHY9612200 and the
Swiss
National Science Foundation.  We acknowledge helpful discussions with
S. {\AA}berg, Y. Alhassid, F. Izrailev, S. Tomsovic, M. Vojta and D.
Weinmann.


\begin{thebibliography}{99}
\bibitem{srs} Due to SRS, states in the tail of each block's MBDOS have
$\sigma = \left| \sigma_z \right|$.
\end{thebibliography}
\end{document}